\documentclass[twocolumn]{aastex63}

\usepackage{lineno}

\usepackage{natbib}
\usepackage{amsmath}
\usepackage{threeparttable}
\usepackage{amssymb}
\usepackage{graphicx}
\usepackage{longtable}
\usepackage{appendix}

\defcitealias{Solomon_1987}{S87}
\defcitealias{Bolatto_2008}{B08}

\shortauthors{Imara, Forbes \& Weaver}

\begin{document}

\title{Touching the Stars: Using High-Resolution 3D Printing to  Visualize Stellar Nurseries}   

\author{Nia Imara}
\affil{University of California, Santa Cruz, 1156 High Street, Santa Cruz, CA 95064}
\email{nimara@ucsc.edu}

\author{John C. Forbes}
\affil{Center for Computational Astrophysics, 162 5th Avenue, New York, NY 10010}

\author{James C. Weaver}
\affil{School of Engineering and Applied Science, Harvard University, Cambridge, MA 02138}

\begin{abstract}
Owing to their intricate variable density architecture, and as a principal site of star formation, molecular clouds represent one of the most functionally significant, yet least understood features of our universe.  To unravel the intrinsic structural complexity of molecular clouds, here we leverage the power of high-resolution bitmap-based 3D printing, which provides the opportunity to visualize astrophysical structures in a way that uniquely taps into the human brain’s ability to recognize patterns suppressed in 2D representations.  Using a new suite of nine simulations, each representing different physical extremes in the turbulent interstellar medium, as our source data, our workflow permits the unambiguous visualization of features in the 3D-printed models, such as quasi-planar structures, that are frequently obscured in traditional renderings and animations.  Our bitmap-based 3D printing approach thus faithfully reproduces the subtle density gradient distribution within molecular clouds in a tangible, intuitive, and visually stunning manner.  While laying the groundwork for the intuitive analysis of other structurally complex astronomical data sets, our 3D-printed models also serve as valuable tools in educational and public outreach endeavors. 
\\
\end{abstract}

\section{Introduction}

As a principal driver of galaxy evolution and the precondition for planets, star formation is one of the most important phenomena in the universe. Because it involves a wide range of physical processes and operates over a broad range of spatial and temporal scales, star formation within molecular clouds is also one of the most poorly understood processes, and questions regarding the details of these phenomena remain largely unresolved \citep{mckee_2007, krumholz_2019}.  Thousands of these vast (20-100 pc), massive ($10^4-10^6 M_\odot$) complexes of gas and dust exist in the Milky Way, and they contain most of the molecular mass in the Galaxy \citep{Rice_2016, Miville-Deschenes_2017}.  Being composed primarily of cold ($\sim$15 K) molecular gas, they are best observed at longer wavelengths.  The earliest radio observations of molecular clouds demonstrated that they exhibit a hierarchical density structure, with clumps of gas at a wide range of size and column/surface density scales \citep[e.g.,][]{Williams_2000}.  Moreover, molecular clouds are threaded by networks of elongated, overdense filaments, which are postulated to play an essential role in the formation of cores and protostars \citep{Molinari_2010, Andre_2014}.  
 
\begin{figure*}[ht!]
    \centering
    \includegraphics[width=\textwidth]{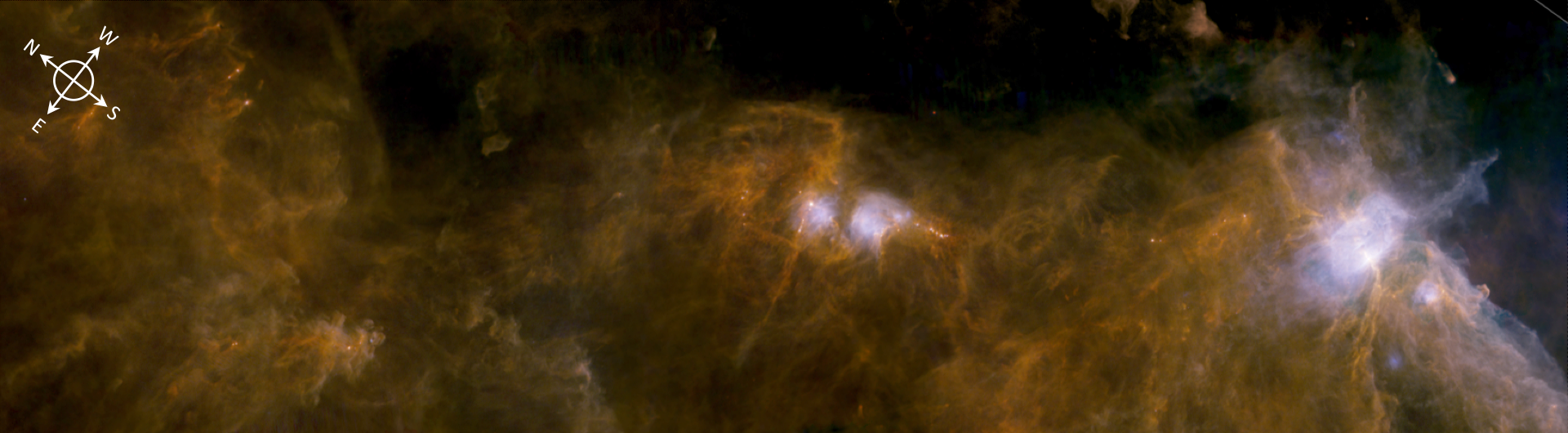}
    \caption{ Far-infrared image of dust emission in the Orion B molecular cloud, from ESA's \textit{Herschel} Space Observatory.  Blue, red, and green correspond to observations at 70 $\mu$m, 160 $\mu$m, and 250 $\mu$m, respectively.  Networks of long, thin filaments thread the cloud, and are sometimes accompanied by bright, compact cores and protostars. Image credit: European Space Agency (ESA).}
    \label{fig:orionb}
\end{figure*}

The clumpy and filamentary structure of molecular clouds has been observed in spectral line emission in radio and sub-millimeter bands, in dust extinction against background starlight at optical and near-infrared wavelengths, and in dust emission at far-infrared and sub-millimeter wavelengths (Figure \ref{fig:orionb}).  A main goal of such observations is to infer the volume density distribution in these molecular clouds.  Quantities that depend on the volume density include the fraction of high-density gas, which correlates with star formation rate \citep[e.g.,][]{Lada_2010}, and the star formation efficiency, both key ingredients in models and simulations attempting to understand the mechanisms of star formation \citep[e.g.,][]{Krumholz_2005}.  But unlike other molecular cloud properties that can be directly inferred from two-dimensional maps, such as surface density and mass, the volume density is fundamentally three-dimensional and is subject to large observational uncertainties arising from simplistic assumptions about the 3D geometry of the observed objects, which likely introduce substantial bias and scatter in the recovered quantities \citep{Hu_2021}.

Simulations have long been used to provide insight into the three-dimensional morphology of stellar nurseries, whose intricate structure bears the imprint of the physics that leads to their formation and shapes their evolution.   Modern simulations are now at a level of sophistication to include self-gravity, turbulence, and magnetic fields, the dominant physical processes that shape cloud environments \citep[e.g.,][]{Vazquez-Semadeni_1994, Klessen_2001, Banerjee_2006,Girichidis_2012, Federrath_2013}.  Even under a range of initial conditions, simulations of the turbulent interstellar medium (ISM) have successfully produced features bearing some resemblance to observed filaments \citep[e.g.,][]{Smith_2014, Kirk_2015, Federrath_2021}.  While there is general agreement that supersonic turbulence leads to filamentary structure and self-gravity results in the further contraction of overdensities, it remains uncertain how this substructure evolves and how it is tied to the formation of stars.
 
Ultimately, our ability to fully explore information available in observed and simulated data is limited by the tools we use to represent them.  Astronomy is fundamentally a visual science and has long used data visualization to drive discovery and communicate new knowledge.  The rendering of 3D astrophysical bodies is as old as cave art, with Huygen’s sketches of Mars and modern simulations of the cosmic web following in the age-old tradition of using visual representations to understand natural phenomena.  As technological advancements have been accompanied by increases in the scale and complexity of astronomical data, visualization of three-dimensional (3D) data has been an area of active innovation  \cite[e.g.,][]{Punzo_2015, Naiman_2017, Kent_2019}.  A variety of software packages have made sophisticated tools readily available to the community, yet even with the most advanced volume rendering techniques, the projection of 3D data onto a 2D surface, results in the loss of information. By approaching data visualization from the standpoint of a personal or shared \textit{experience}, recent efforts have attempted to unite astronomy with virtual reality and augmented reality \citep[e.g.,][]{Vogt_2013, Arcand_2018, Orlando_2019}.  
 
The advent of 3D printing provides the opportunity to represent astrophysical structures in a way that more fully taps into the human brain’s ability in pattern recognition.  Moreover, inherently interactive 3D structures can engage our intuition in ways that 2D representations cannot. In art, the purpose of sculpture is not to replace painting, nor is sculpture an objectively superior art form.  Likewise, 3D printing should be viewed as a complementary approach to other graphical means of data exploration, one having the potential to illuminate otherwise concealed patterns and features.

While 3D printing has been used previously for the visualization of astronomical data sets, including stellar winds \citep{Madura_2015}, the cosmic web \citep{Diemer_2017}, and the cosmic microwave background \citep{Clements_2017},  in all of these efforts, complex 3D structures of variable density are reduced to single surfaces, and thus exclude information beyond the arbitrarily designated isosurface boundary \citep{hosny_2018}.  In contrast, our multi-material bitmap-based 3D printing approach presented here (see Figure \ref{fig:workflow}) preserves all of the intrinsic density variation present in the native astronomical source files, thus resulting in more detailed and faithful representations of point cloud-based volumetric data, and thus increases the instructive value of the resulting tangible models.

In this paper, we present the results of the first 3D printed stellar nurseries, amongst the most physically complex structures in the cosmos. Through these efforts, our overarching goal is to gain insight into how gravity, turbulence, and magnetic fields shape the sub-structure of molecular clouds.  To that end, we ran a suite of simulations of the turbulent ISM designed to isolate and emphasize the roles of these various processes.

\section{Methods}\label{sec:methods}

Our overall goal is to investigate how the three fundamental physical processes governing the evolution of the ISM---turbulence, gravity, and magnetic fields---determine the morphology of sub-structures intimately related to star formation.   To isolate their particular impacts on the development of sub-structures---including cores, filaments, and sheets---we set up a suite of simulations representing different physical extremes.   The simulations are then used as source data for our 3D-printed molecular clouds, representing the following physical scenarios: (1) A fiducial model representing ``normal" physical parameters typically observed in real molecular clouds; (2) low Mach number; (3) high Mach number; (4) low Alfven Mach number; (5) high Alfven Mach number; (6) low virial parameter; (7) high virial parameter; (8) purely solenoidal turbulence; and (9) purely compressive turbulence.

\subsection{Simulations} 

The simulations presented here are run with version 2.5 of the ENZO adaptive mesh refinement (AMR) code \citep{Bryan_2014, Brummel-Smith2019}, though for this particular problem we do not use AMR, since the entire computational domain contains structure that we would like to resolve. We use the Monotone Upstream-centered Schemes for Conservation Laws (MUSCL) hydrodynamics solver \citep{WangAndAbel_2009}, a second-order Godunov method with second-order Runge-Kutta time integration and a Dedner method \citep{Dedner_2002}  for maintaining $\nabla \cdot \vec{B} = 0$ to a few percent. The Riemann problem is solved with the Harten-Lax-van Leer two-wave solver \citep{toro_1997} and a piecewise-linear reconstruction method.  

The evolution equations can be rewritten in terms of the dimensionless ratios $\mathcal{M}=v/c_s$, the Mach number; $\mathcal{M_A}=v/v_A$, the Alv\'enic Mach number; and $\alpha_\mathrm{vir} = 5 v^2 / (3 G \rho L^2)$, the virial parameter. Here $c_s$ is the sound speed, $v$ is the typical velocity of the gas, $v_\mathcal{A}=B/\sqrt{4\pi\rho}$ is the Alfv\'en speed, $\rho$ is the gas density, $L$ is the box size, and $B$ is the magnitude of the magnetic field. The evolution equations are identical if $\rho$, $L$ and $B$ are simultaneously scaled by $x$, $x^{-1/2}$, and $x^{1/2}$ respectively, where $x$ is an arbitrary positive number \citep{McKee_2010, Krumholz_2011}, so we choose fiducial values of these dimensionless ratios, and vary each up or down separately, while keeping their range reasonable for giant molecular clouds (GMCs). This produces 7 sets of parameters, plus 2 from varying the solenoidal vs. compressive weighting of the driving force (see Table in Figure \ref{fig:workflow}). While the table lists physical units, their overall scale is arbitrary and they may be rescaled with the transformation above.

The turbulent driving in our simulations closely follows \citep[][ProblemType = 59 in ENZO]{Schmidt_2009}, in which the driving force per unit mass at all locations in the simulation undergoes a damped random (DrivenFlowSeed=842091) walk in $k$-space, characterized by an auto-correlation timescale, a profile in $k$-space, and a splitting of the driving into solenoidal and compressive modes. We set the auto-correlation timescale to be a box crossing time $L/v$ (DrivenFlowAutoCorrl=1), the $k$-space profile to be parabolic (DrivenFlowProfile=2) centered on a length-scale equal to half (DrivenFlowAlpha=2) the box size (see \citet{Bialy_2020} for simulations that vary this scale) with a characteristic bandwidth equal to the location of the peak of the driving (DrivenFlowBandWidth=1). The box has periodic boundary conditions, and is initialized with zero velocity, a uniform density, and a uniform magnetic field pointing in the $x$-direction. The simulations are evolved for three crossing times before self-gravity is turned on. Because the simulations do not include a mechanism to resolve gravitational collapse to small scales, e.g., sink particles, artificial jeans support, or feedback, the simulations quickly develop cells that violate the Truelove criterion \citep{Truelove_1997, Federrath_2011}, so we only follow the collapse for 0.1 box crossing times. Dense clumps present in the simulation when self-gravity is turned on will of course collapse much faster than this, limiting the time span over which the solution is expected to be reliable, i.e., free of artificial fragmentation. 
\\

\subsection{3D Printing}
\begin{figure*}[t!]
    \centering
    \includegraphics[width=150mm]{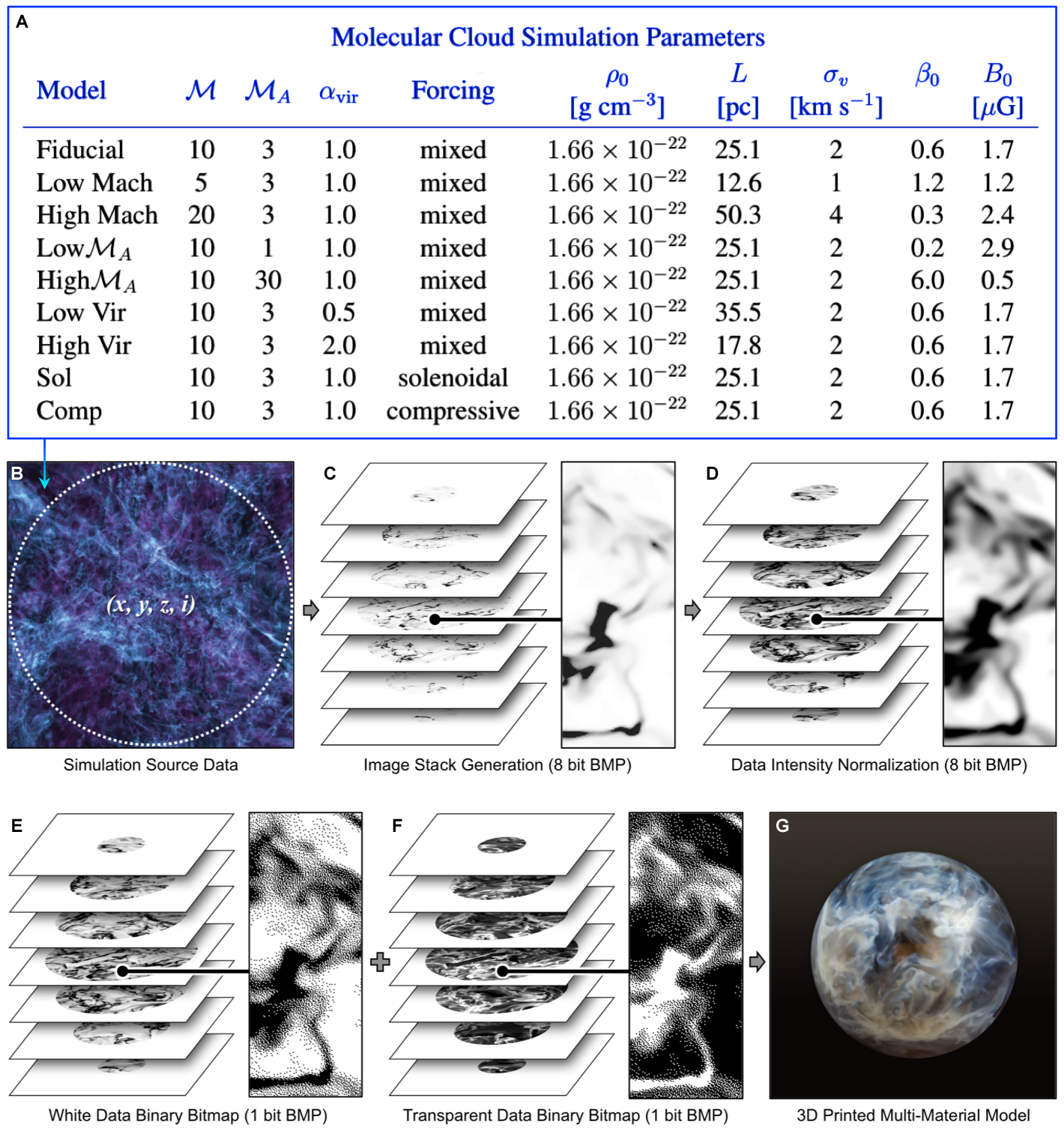}
    \caption{Multi-material additive manufacturing workflow for the production of tangible (physical) models from astronomical simulation data sets.  The molecular cloud simulation parameters (A) are used to construct a point cloud-based 3D data set (B), which is cropped into a spherical volume (denoted by the dotted circle) and sliced into a stack of 8-bit BMP files that correspond to the \textit{x}, \textit{y}, and \textit{z} resolution of the 3D printing platform (C).  The bitmap slice intensity values are normalized in order to effectively visualize the features of interest (D) and then, via a diffusion dithering step, are separated into their corresponding positive (E) and negative (F) 1-bit bitmaps for 3D printing (which are shown at higher magnification in the right panels).  The resulting 3D-printed sphere, which was produced from transparent and white photopolymers is shown in (G).}
    \label{fig:workflow}
\end{figure*}

\begin{figure*}[ht!]
    \centering
    \includegraphics[width=1\textwidth]{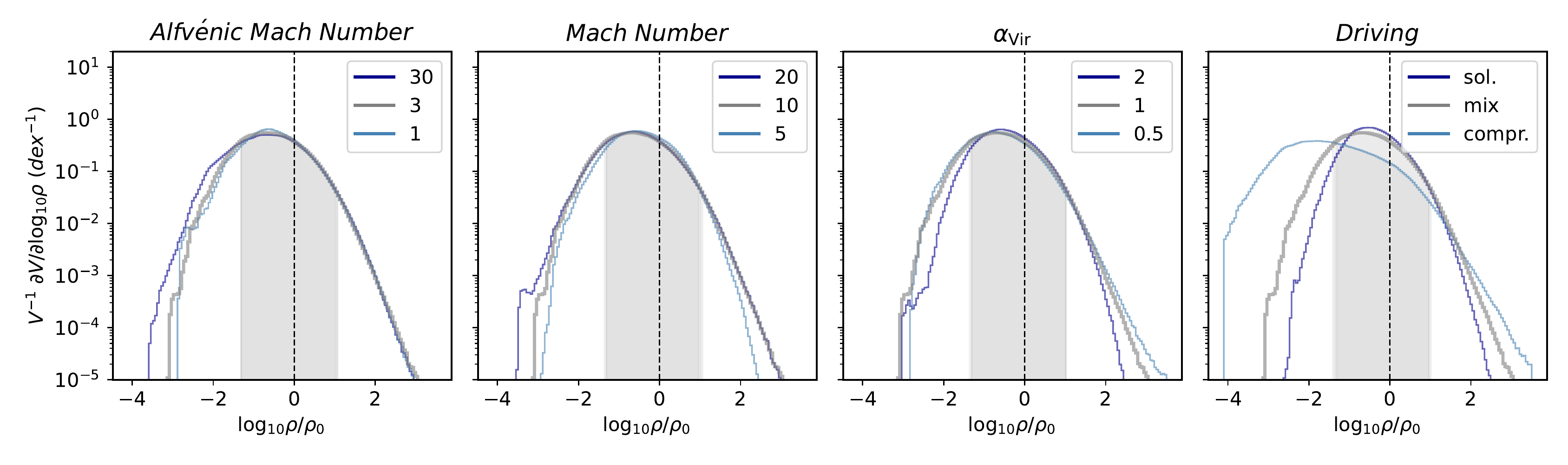}
    \caption{Simulation density distribution per unit volume. Each panel compares the fiducial case (in gray) to a high and low value of each varied parameter. The truncated light gray shaded regions in each panel correspond to the the data range that best emphasizes the structural details surrounding the centers of star formation, which was the focus of the 3D printing studies. 
    }
    \label{fig:pdfs}
\end{figure*}

The methodology employed in the present study focused on establishing format compatibility between the data outputs from our astronomical simulations and the inputs required for multi-material UV-crosslinkable photopolymer 3D printing processes \citep{bader_2016, bader_2018, hosny_2018}.

In this multi-step approach (Figure \ref{fig:workflow}), the simulation source data (Figure \ref{fig:workflow}A) was first truncated (denoted by the shaded regions in Figure 3) in order to specifically emphasize the complex environments that produce the star-forming high density cores.  The resulting volumetric renderings (Figure \ref{fig:workflow}B) were then exported as a set of 8 bit grayscale bitmaps at the desired printer resolution and slice thickness. To delineate the outer contours of the 3D-printed object, a set of masks corresponding to the intended spherical form were subtracted from the imaging data (Figure \ref{fig:workflow}C). Using 3D printer platform-specific look-up tables developed by \cite{bader_2018}, the image slice intensity values were normalized in order to reveal the structural elements of interest (Figure \ref{fig:workflow}D). Finally, the images were converted into binary bitmaps of black [1] and white [0] pixels using the Floyd–Steinberg dithering algorithm \citep{Floyd_1976}. This process results in the integrated density ratio of black to white pixels in the new images, approximating the grayscale values in the original source files, conceptually similar to the use of halftone image processing techniques in newsprint. When 3D printing with two materials (in this case, opaque and transparent photopolymers), two bitmap stacks must be created, one for each material, with one stack (Figure \ref{fig:workflow}E) being the inverse of the other (Figure \ref{fig:workflow}F). The multiple data transformation steps (Figure \ref{fig:workflow}C-F) were performed simultaneously on the entire stack of slices in a sequential fashion through use of the batch processing function in Adobe Photoshop (Adobe Inc., Mountain View, CA). As with previous efforts aimed at the 3D printing of volumetric data sets that exhibit smooth gradients in signal intensity \citep{hosny_2018}, this bitmap based workflow entirely bypasses the conventional data thresholding and STL file generation-based approaches that can over- or under-estimate feature sizes or create object boundary designations lacking in the original source data. 

The resulting bitmaps were printed on a Connex500 (Stratasys, Eden Prairie, Minnesota) multi-material 3D printer from clear (RGD810) and white (RGD835) photopolymers. The spheres (Figure \ref{fig:workflow}F) were mechanically polished in a three step process which included two high energy centrifugal disc finishers and a low energy vibratory polisher (Bel Air Finishing, North Kingstown, Rhode Island), and were photographed with a Nikon D7100 DSLR camera equipped with a Nikkor 50 mm 1.4 lens using either reflective or transmitted LED light sources.

\begin{figure*}[ht!]
    \centering
    \includegraphics[width=160mm]{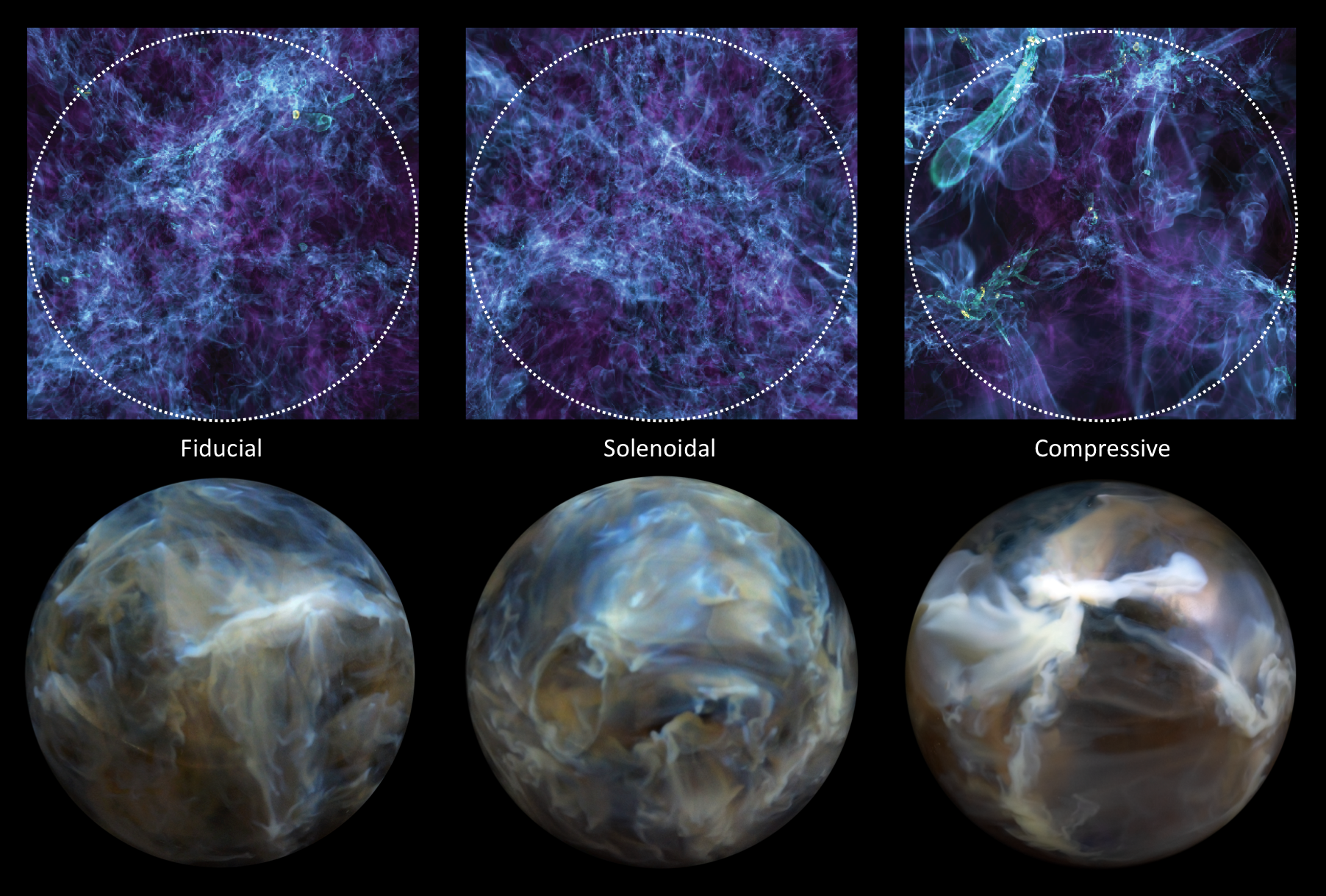}
    \caption{Comparisons of 2D volume renderings of a subset of our simulations (upper row) and their corresponding 3D-printed counterparts (lower row).  The dotted circle in each case denotes the border of the cropped spherical volume.  From left to right, the fiducial model, purely solenoidal turbulence, and purely compressive turbulence. For scale, each 3D-printed sphere measures 8 cm in diameter.}
    \label{fig:simulations}
\end{figure*}

\section{Results}\label{sec:clouds}

\begin{figure*}[ht!]
    \centering
    \includegraphics[width=\textwidth]{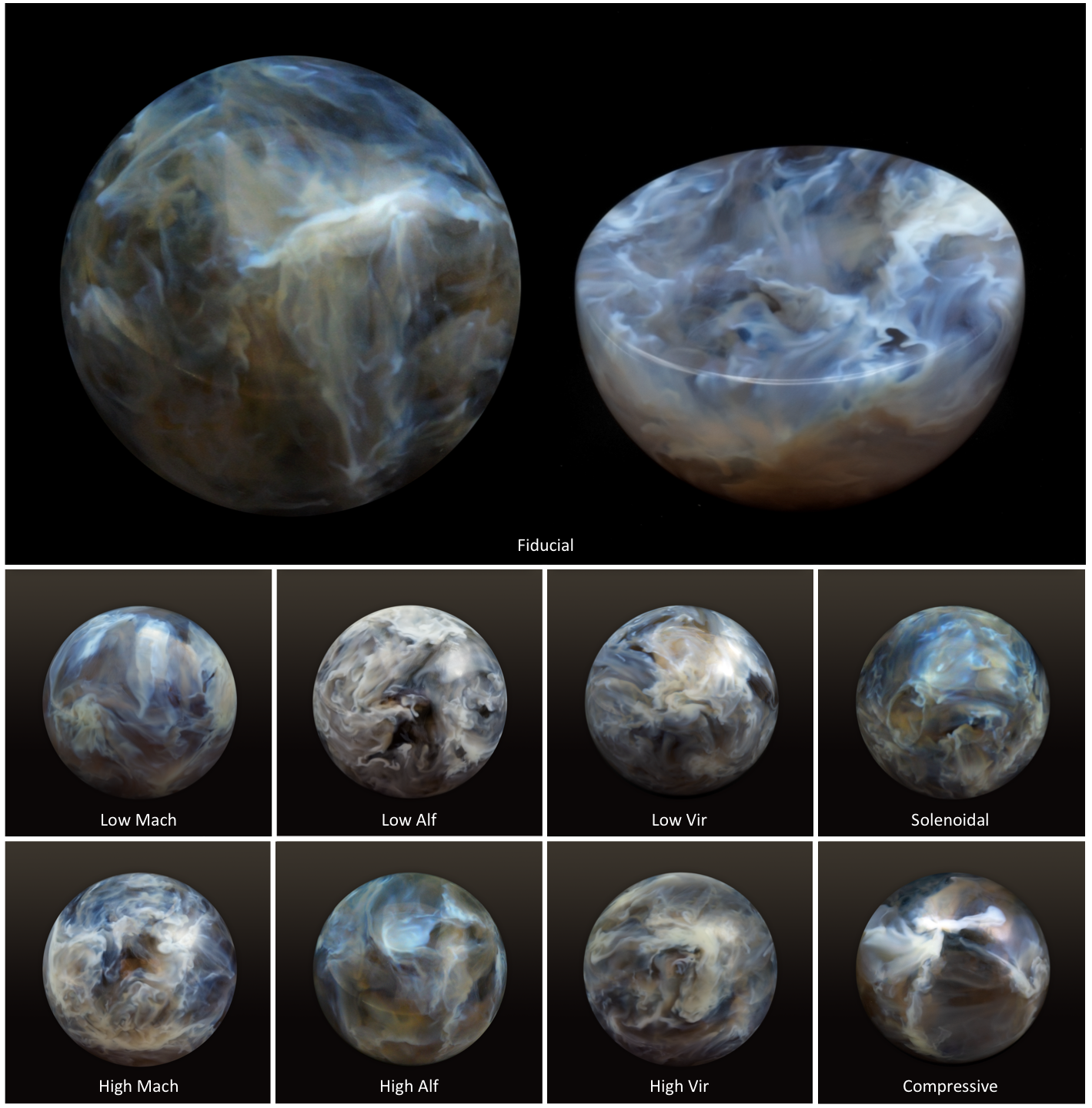}
    \caption{Photographs of 3D-printed molecular clouds.  Each of the 8 cm diameter spheres represent the full extent of the simulated cuboidal volume (see Figures 2 and 3), with the eight lower conditions compared to the fiducial control (shown as both a full sphere, and a bisected sphere to reveal the mid-plane data) in the upper panel.  Lighter material corresponds to regions of higher density, while darker areas represent regions of low density and voids.}
    \label{fig:spheres}
\end{figure*}

\begin{figure*}
\centering
 \includegraphics[width = 6in]{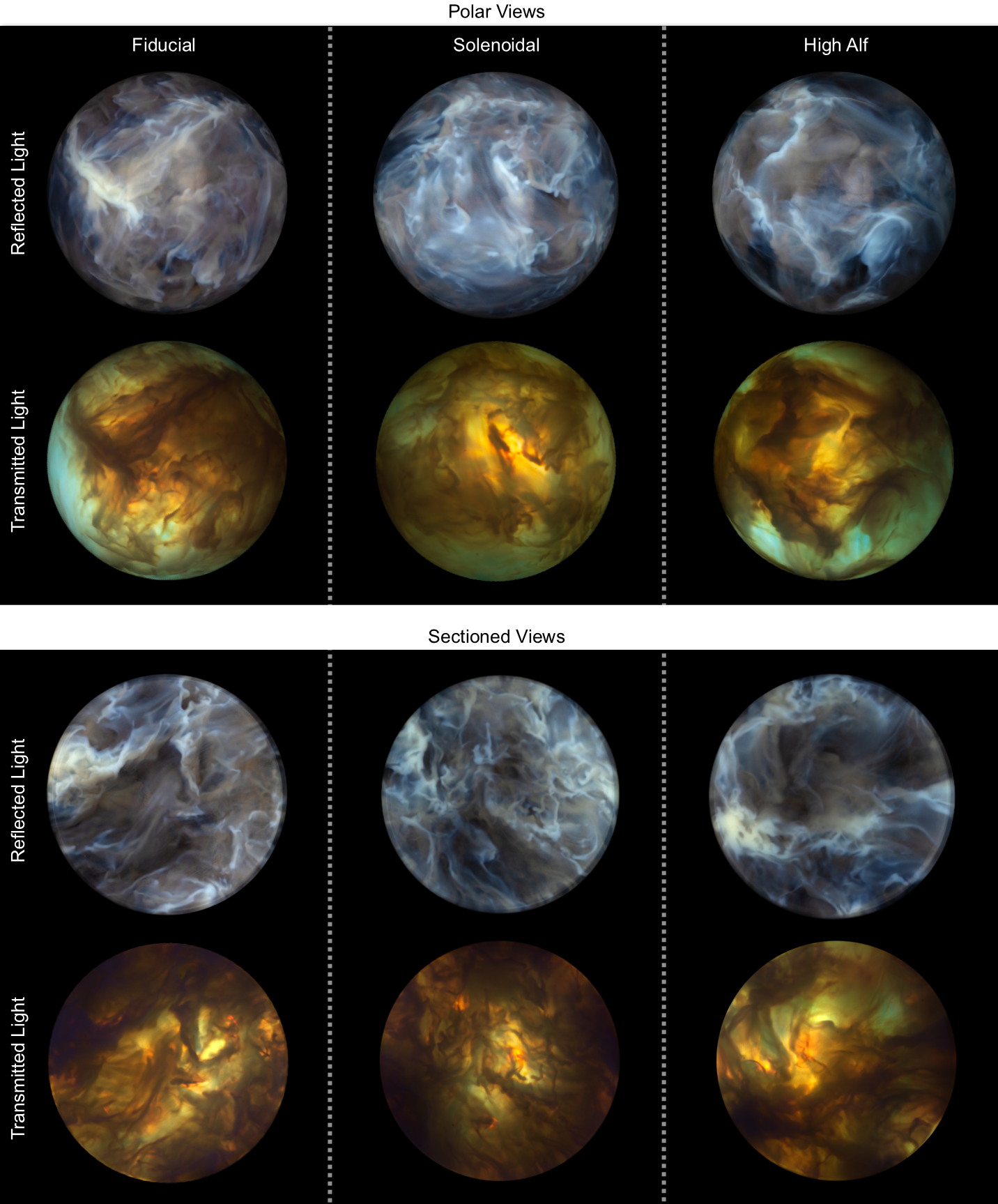}
    \caption{Correlative imaging of 3D printed half spheres under different illumination regimes.  From left to right, the fiducial, purely solenoidal turbulence, and High Alfven Mach number models.  Each of the image sets correspond to either polar views (upper set) or mid-plane/sectioned views (lower set), and show the two orientations when viewed with either reflected or transmitted (back-lit) illumination.  The yellows and oranges created from the back-lit illumination are due to the density-dependent differential light scattering from the titanium dioxide nanoparticles that were used to formulate the white opaque resin component of these models.}
    \label{fig:half_spheres}
\end{figure*}

Figure \ref{fig:simulations} shows comparisons between volume renderings of a subset of our simulations produced with yt \citep{Turk2011} using a transfer function constructed with a handful of narrow gaussians centered at different log-densities, and their corresponding 3D-printed counterparts.  Figure  \ref{fig:spheres} displays photographs of all nine 3D-printed molecular clouds, each having a diameter of 8 cm, and each representing a different simulation.  For aesthetic reasons, we printed the clouds as spheres, but we could have chosen any arbitrary geometry.  We also printed three half-spheres for the fiducial, purely solenoidal turbulence, and high Alfv\'en Mach number models (see Figure \ref{fig:half_spheres}).  In the photographs with reflected lighting, greater concentrations of white material represent regions of higher density.  The grayer, darker regions represent regions of low density and voids.  We also photographed the half-spheres with back-lighting.  In these photos, regions of high-density (i.e., ``high-extinction" material) appear opaque, while regions of low-density (where the light penetrates the sphere) appear various shades of yellows, oranges, and reds.

Overall, the wispy, filamentary, and generally ``cloud-like" appearance of the spheres is reminiscent of traditional visualizations of both turbulent box simulations and observations of the ISM.  What becomes immediately apparent upon closer examination are three striking characteristics that are not readily apparent, or are entirely imperceptible, in conventional volume renderings. 

First, the sub-structure in individual spheres is generally very continuous.  In other words, individual filaments or ``wisps" can extend over much longer distances than one might anticipate from 2D images of simulations or observations.  In any one of the spheres, one can observe a prominent filament near the surface that winds its way toward the interior of the cloud, bending and looping around, until it disappears from view.  As such a filament twists its way through the cloud, its average density per unit length may gradually vary, but the overall continuity of the filament is preserved, and it is clear that the structure is coherent.  In a 2D image, such a structure would appear truncated and disconnected from other filaments in the environment.

Second, the spheres reveal the presence of complex, sheet-like structures that are difficult to perceive in traditional 2D volume renderings.  On flat surfaces, and even in animations, it is a challenge to discern the continuity of sheets, which, depending on the viewing angle or projection, may be mistaken for filaments.

Third, the overall qualitative differences between spheres is noteworthy.  Such differences have been observed and quantified previously, but the visual manifestation of different physical conditions has a much stronger cognitive impact, and is perhaps more intuitive, in the 3D prints. In the low- and high-$\mathcal{M_A}$ spheres, for example, the effects of the magnetic field on cloud morphology are accentuated in an immediate way. While in the low-$\mathcal{M_A}$ sphere the larger magnetic field acts to suspend material into sharply twisting structures, in the high-$\mathcal{M_A}$ sphere, structures are generally more collapsed leaving large voids. The high-Mach and low-$\mathcal{M_A}$ spheres are similarly ``crowded" with suspended structures, a result of these two models having greater support against gravitational collapse away from the densest structures. Yet the sub-structure of the Low-$\mathcal{M_A}$ sphere is generally more compact, with tighter curves and twists---the consequence of having a smaller velocity dispersion than in the high-Mach case.

\section{Discussion}\label{sec:discussion}
There are several qualitative stories used to explain the structures seen in simulations of supersonic turbulence. Without gravity or magnetic fields, a random point in a turbulent box will be subject to the repeated passage of shocks and rarefactions. For isothermal turbulence, the shock jump conditions yield a density increase by a factor proportional to the Mach number squared. In steady state, this will produce the lognormal probability density functions (PDFs) measured in such simulations, with a characteristic width closely related to the Mach number \citep[e.g.,][]{Padoan_1997, Padoan_2002, Federrath_2008}. This picture of a shock-dominated medium is further confirmed  by identifying individual shocks and their profiles along the direction of motion of the shock \citep{Robertson_2018}.

With the inclusion of self-gravity, a power-law tail is observed to develop at the high-density end of the PDF (Figure \ref{fig:pdfs}), which is understood as the result of individual structures collapsing under their own self-gravity on the local freefall time \citep[e.g.,][]{Ballesteros-Paredes_2011, Federrath_2013}. This process is expected to bear some resemblance to the collapse of cosmological structures, where on large scales the density field should initially collapse in planar, then filamentary, then point-like structures \cite{zeldovich_1970}. The similarity between cosmological and turbulent self-gravitating collapse has led to some exploration of the excursion set formalism \citep{press_1974} to model the collapse of cores and hence the initial mass function and star formation rate \citep[e.g.,][]{hennebelle_2008, hopkins_2012}.

Strong magnetic fields are also expected to produce density structures in turbulence. If gas is constrained to follow magnetic field lines, as is the case when the magnetic pressure is strong compared to the thermal pressure \citep{Beattie_2021}, filaments aligned with the large-scale field are produced, which is the favored explanation for filaments observed in the low-density warm neutral medium. Each qualitative picture is expected to be relevant in different regimes, according to the values of the dimensionless numbers describing the problem. 

It is likely that the individual nonlinear structures in the density field, i.e., the filaments and pancakes that form just as in the cosmological case, will not substantially affect the resulting core or stellar mass function \citep[e.g.,][]{hopkins_2013}. However, empirically galaxies are affected by their proximity to cosmological structures, with galaxies in closer proximity having suppressed star formation rates \citep[e.g.,][]{winkel_2021}, likely the result of adiabatic heating associated as gas follows the dark matter into these structures \citep{peng_2010}. Since this effect is presumably much weaker in the near-isothermal conditions of GMCs, another correlation observed in the cosmological case may be more relevant, namely the apparent relationship between galaxies' spins and the orientation of the structure in which they reside  \citep[e.g.,][]{an_2021}. The prominence of these structures in our simulated GMCs may explain the astroseismically-inferred alignment of stellar spins within open clusters \citep{corsaro_2017} without relying on initial rapid large-scale rotation of the cloud. Instead, we suggest that this alignment could occur via pancakes and filaments, in agreement with simulations that suggest that stellar spins are likely determined from local structures around the cores \citep[e.g.,][]{kuznetsova_2019}. The prominence of these structures may also play a substantial role in understanding how to generate realistic initial conditions for N-body stellar cluster simulations, which currently must rely on simple approximate initializations to produce statistical samples of cluster evolution  \citep[e.g.,][]{torniamenti_2021}.

\section{Conclusion}\label{sec:conclusion}
As demonstrated in the present study, our bitmap-based 3D printing approach faithfully reproduces the subtle density gradient distribution within molecular clouds in a tangible, intuitive, and visually stunning form factor. While the current 3D-printed models were only fabricated using two photopolymers (white and clear), recent advances in the development of full-color and optically transparent bitmap-based inkjet-based 3D printing technologies  \citep{bader_2018} permit the future production of full-color molecular cloud data sets that include additional kinematic properties, such as velocity. One of the intriguing observations from the 3D-printed half-spheres is that they permit the non-distorted visualization of a specific data plane within the 3D printed volume.  This observation demonstrates the full potential of combining curved and flat surfaces in the 3D-printed models for highlighting different features of interest, while also maintaining a standardizable viewing orientation for direct data visualization correlations under different lighting regimes (Figure \ref{fig:half_spheres}).  Beyond the production of 3D-printed simulations, we can also leverage these approaches for the production of tangible models based on observational data of local molecular clouds, such as Orion B (Figure \ref{fig:orionb}), with a primary objective of determining the efficacy of using line-of-sight velocity to define distinct sub-structures from these observations.  We expect that these efforts will help elucidate the relative coherency of sub-structures that, as they evolve, undergo fragmentation, and form stars.

In addition to laying groundwork for the intuitive analysis of other structurally complex astronomical data sets, our 3D printed models serve as valuable tools in educational and public outreach endeavors.

\acknowledgements
We thank Saurabh Mhatre for assistance with the model photography.  We thank an anonymous referee for their insightful questions and feedback.  JCF thanks the Flatiron Institute, a division of the Simons Foundation, for funding support for this project.


\bibliography{paper_arxiv}

\begin{thebibliography}{}
\expandafter\ifx\csname natexlab\endcsname\relax\def\natexlab#1{#1}\fi
\providecommand{\url}[1]{\href{#1}{#1}}
\providecommand{\dodoi}[1]{doi:~\href{http://doi.org/#1}{\nolinkurl{#1}}}
\providecommand{\doeprint}[1]{\href{http://ascl.net/#1}{\nolinkurl{http://ascl.net/#1}}}
\providecommand{\doarXiv}[1]{\href{https://arxiv.org/abs/#1}{\nolinkurl{https://arxiv.org/abs/#1}}}

\bibitem[{{An} {et~al.}(2021){An}, {Kim}, {Moon}, \& {Yoon}}]{an_2021}
{An}, S.-H., {Kim}, J., {Moon}, J.-S., \& {Yoon}, S.-J. 2021, arXiv e-prints,
  arXiv:2105.12741.
\newblock \doarXiv{2105.12741}

\bibitem[{{Andr{\'e}} {et~al.}(2014){Andr{\'e}}, {Di Francesco},
  {Ward-Thompson}, {Inutsuka}, {Pudritz}, \& {Pineda}}]{Andre_2014}
{Andr{\'e}}, P., {Di Francesco}, J., {Ward-Thompson}, D., {et~al.} 2014, in
  Protostars and Planets VI, ed. H.~{Beuther}, R.~S. {Klessen}, C.~P.
  {Dullemond}, \& T.~{Henning}, 27,
  \dodoi{10.2458/azu\_uapress\_9780816531240-ch002}

\bibitem[{{Aracand} {et~al.}(2018){Aracand}, {Jiang}, {Price}, {Watzke},
  {Sgouros}, \& {Edmonds}}]{Arcand_2018}
{Aracand}, K.~K., {Jiang}, E., {Price}, S., {et~al.} 2018, Communicating
  Astronomy with the Public Journal, 24, 17–24

\bibitem[{Bader {et~al.}(2016)Bader, Kolb, Weaver, \& Oxman}]{bader_2016}
Bader, C., Kolb, D., Weaver, J.~C., \& Oxman, N. 2016, 3D Printing and Additive
  Manufacturing, 3, 71, \dodoi{10.1089/3dp.2016.0026}

\bibitem[{{Bader} {et~al.}(2018){Bader}, {Kolb}, {Weaver}, {Sharma}, {Hosny},
  {Costa}, \& {Oxman}}]{bader_2018}
{Bader}, C., {Kolb}, D., {Weaver}, J.~C., {et~al.} 2018, Science Advances, 4,
  eaas8652, \dodoi{10.1126/sciadv.aas8652}

\bibitem[{{Ballesteros-Paredes} {et~al.}(2011){Ballesteros-Paredes},
  {V{\'a}zquez-Semadeni}, {Gazol}, {Hartmann}, {Heitsch}, \&
  {Col{\'\i}n}}]{Ballesteros-Paredes_2011}
{Ballesteros-Paredes}, J., {V{\'a}zquez-Semadeni}, E., {Gazol}, A., {et~al.}
  2011, \mnras, 416, 1436, \dodoi{10.1111/j.1365-2966.2011.19141.x}

\bibitem[{{Banerjee} {et~al.}(2006){Banerjee}, {Pudritz}, \&
  {Anderson}}]{Banerjee_2006}
{Banerjee}, R., {Pudritz}, R.~E., \& {Anderson}, D.~W. 2006, \mnras, 373, 1091,
  \dodoi{10.1111/j.1365-2966.2006.11089.x}

\bibitem[{{Beattie} {et~al.}(2021){Beattie}, {Mocz}, {Federrath}, \&
  {Klessen}}]{Beattie_2021}
{Beattie}, J.~R., {Mocz}, P., {Federrath}, C., \& {Klessen}, R.~S. 2021,
  \mnras, \dodoi{10.1093/mnras/stab1037}

\bibitem[{{Bialy} \& {Burkhart}(2020)}]{Bialy_2020}
{Bialy}, S., \& {Burkhart}, B. 2020, \apjl, 894, L2,
  \dodoi{10.3847/2041-8213/ab8a32}

\bibitem[{Brummel-Smith {et~al.}(2019)Brummel-Smith, Bryan, Butsky, Corlies,
  Emerick, Forbes, Fujimoto, Goldbaum, Grete, Hummels, hoon Kim, Koh, Li, Li,
  Li, OShea, Peeples, Regan, Salem, Schmidt, Simpson, Smith, Tumlinson, Turk,
  Wise, Abel, Bordner, Cen, Collins, Crosby, Edelmann, Hahn, Harkness,
  Harper-Clark, Kong, Kritsuk, Kuhlen, Larrue, Lee, Meece, Norman, Oishi,
  Paschos, Peruta, Razoumov, Reynolds, Silvia, Skillman, Skory, So, Tasker,
  Wagner, Wang, Xu, \& Zhao}]{Brummel-Smith2019}
Brummel-Smith, C., Bryan, G., Butsky, I., {et~al.} 2019, Journal of Open Source
  Software, 4, 1636, \dodoi{10.21105/joss.01636}

\bibitem[{{Bryan} {et~al.}(2014){Bryan}, {Norman}, {O'Shea}, {Abel}, {Wise},
  {Turk}, {Reynolds}, {Collins}, {Wang}, {Skillman}, {Smith}, {Harkness},
  {Bordner}, {Kim}, {Kuhlen}, {Xu}, {Goldbaum}, {Hummels}, {Kritsuk}, {Tasker},
  {Skory}, {Simpson}, {Hahn}, {Oishi}, {So}, {Zhao}, {Cen}, {Li}, \& {Enzo
  Collaboration}}]{Bryan_2014}
{Bryan}, G.~L., {Norman}, M.~L., {O'Shea}, B.~W., {et~al.} 2014, \apjs, 211,
  19, \dodoi{10.1088/0067-0049/211/2/19}

\bibitem[{{Clements} {et~al.}(2017){Clements}, {Sato}, \& {Portela
  Fonseca}}]{Clements_2017}
{Clements}, D.~L., {Sato}, S., \& {Portela Fonseca}, A. 2017, European Journal
  of Physics, 38, 015601, \dodoi{10.1088/0143-0807/38/1/015601}

\bibitem[{{Corsaro} {et~al.}(2017){Corsaro}, {Lee}, {Garc{\'\i}a},
  {Hennebelle}, {Mathur}, {Beck}, {Mathis}, {Stello}, \&
  {Bouvier}}]{corsaro_2017}
{Corsaro}, E., {Lee}, Y.-N., {Garc{\'\i}a}, R.~A., {et~al.} 2017, Nature
  Astronomy, 1, 0064, \dodoi{10.1038/s41550-017-0064}

\bibitem[{{Dedner} {et~al.}(2002){Dedner}, {Kemm}, {Kr{\"o}ner}, {Munz},
  {Schnitzer}, \& {Wesenberg}}]{Dedner_2002}
{Dedner}, A., {Kemm}, F., {Kr{\"o}ner}, D., {et~al.} 2002, Journal of
  Computational Physics, 175, 645, \dodoi{10.1006/jcph.2001.6961}

\bibitem[{{Diemer} \& {Facio}(2017)}]{Diemer_2017}
{Diemer}, B., \& {Facio}, I. 2017, \pasp, 129, 058013,
  \dodoi{10.1088/1538-3873/aa6a46}

\bibitem[{{Federrath} \& {Klessen}(2013)}]{Federrath_2013}
{Federrath}, C., \& {Klessen}, R.~S. 2013, \apj, 763, 51,
  \dodoi{10.1088/0004-637X/763/1/51}

\bibitem[{{Federrath} {et~al.}(2021){Federrath}, {Klessen}, {Iapichino}, \&
  {Beattie}}]{Federrath_2021}
{Federrath}, C., {Klessen}, R.~S., {Iapichino}, L., \& {Beattie}, J.~R. 2021,
  Nature Astronomy, 5, 365, \dodoi{10.1038/s41550-020-01282-z}

\bibitem[{{Federrath} {et~al.}(2008){Federrath}, {Klessen}, \&
  {Schmidt}}]{Federrath_2008}
{Federrath}, C., {Klessen}, R.~S., \& {Schmidt}, W. 2008, \apjl, 688, L79,
  \dodoi{10.1086/595280}

\bibitem[{{Federrath} {et~al.}(2011){Federrath}, {Sur}, {Schleicher},
  {Banerjee}, \& {Klessen}}]{Federrath_2011}
{Federrath}, C., {Sur}, S., {Schleicher}, D. R.~G., {Banerjee}, R., \&
  {Klessen}, R.~S. 2011, \apj, 731, 62, \dodoi{10.1088/0004-637X/731/1/62}

\bibitem[{{Floyd} \& {Steinberg}(1976)}]{Floyd_1976}
{Floyd}, R., \& {Steinberg}, L. 1976, Proceedings of the Society of Information
  Display 17, 75

\bibitem[{{Girichidis} {et~al.}(2012){Girichidis}, {Federrath}, {Allison},
  {Banerjee}, \& {Klessen}}]{Girichidis_2012}
{Girichidis}, P., {Federrath}, C., {Allison}, R., {Banerjee}, R., \& {Klessen},
  R.~S. 2012, \mnras, 420, 3264, \dodoi{10.1111/j.1365-2966.2011.20250.x}

\bibitem[{{Hennebelle} \& {Chabrier}(2008)}]{hennebelle_2008}
{Hennebelle}, P., \& {Chabrier}, G. 2008, \apj, 684, 395,
  \dodoi{10.1086/589916}

\bibitem[{{Hopkins}(2012)}]{hopkins_2012}
{Hopkins}, P.~F. 2012, \mnras, 423, 2037,
  \dodoi{10.1111/j.1365-2966.2012.20731.x}

\bibitem[{{Hopkins}(2013)}]{hopkins_2013}
---. 2013, \mnras, 430, 1653, \dodoi{10.1093/mnras/sts704}

\bibitem[{Hosny {et~al.}(2018)Hosny, Keating, Dilley, Ripley, Kelil, Pieper,
  Kolb, Bader, Pobloth, Griffin, Nezafat, Duda, Chiocca, Stone, Michaelson,
  Dean, Oxman, \& Weaver}]{hosny_2018}
Hosny, A., Keating, S.~J., Dilley, J.~D., {et~al.} 2018, 3D Printing and
  Additive Manufacturing, 5, 103, \dodoi{10.1089/3dp.2017.0140}

\bibitem[{{Hu} {et~al.}(2021){Hu}, {Krumholz}, {Federrath}, {Pokhrel}, \&
  {Gutermuth}}]{Hu_2021}
{Hu}, Z., {Krumholz}, M.~R., {Federrath}, C., {Pokhrel}, R., \& {Gutermuth},
  R.~A. 2021, \mnras, 502, 5997, \dodoi{10.1093/mnras/stab356}

\bibitem[{{Kent}(2019)}]{Kent_2019}
{Kent}, B.~R. 2019, in Astronomical Society of the Pacific Conference Series,
  Vol. 523, Astronomical Data Analysis Software and Systems XXVII, ed. P.~J.
  {Teuben}, M.~W. {Pound}, B.~A. {Thomas}, \& E.~M. {Warner}, 3

\bibitem[{{Kirk} {et~al.}(2015){Kirk}, {Klassen}, {Pudritz}, \&
  {Pillsworth}}]{Kirk_2015}
{Kirk}, H., {Klassen}, M., {Pudritz}, R., \& {Pillsworth}, S. 2015, \apj, 802,
  75, \dodoi{10.1088/0004-637X/802/2/75}

\bibitem[{{Klessen} \& {Burkert}(2001)}]{Klessen_2001}
{Klessen}, R.~S., \& {Burkert}, A. 2001, \apj, 549, 386, \dodoi{10.1086/319053}

\bibitem[{{Krumholz}(2011)}]{Krumholz_2011}
{Krumholz}, M.~R. 2011, \apj, 743, 110, \dodoi{10.1088/0004-637X/743/2/110}

\bibitem[{{Krumholz} \& {McKee}(2005)}]{Krumholz_2005}
{Krumholz}, M.~R., \& {McKee}, C.~F. 2005, \apj, 630, 250,
  \dodoi{10.1086/431734}

\bibitem[{{Krumholz} {et~al.}(2019){Krumholz}, {McKee}, \&
  {Bland-Hawthorn}}]{krumholz_2019}
{Krumholz}, M.~R., {McKee}, C.~F., \& {Bland-Hawthorn}, J. 2019, \araa, 57,
  227, \dodoi{10.1146/annurev-astro-091918-104430}

\bibitem[{{Kuznetsova} {et~al.}(2019){Kuznetsova}, {Hartmann}, \&
  {Heitsch}}]{kuznetsova_2019}
{Kuznetsova}, A., {Hartmann}, L., \& {Heitsch}, F. 2019, \apj, 876, 33,
  \dodoi{10.3847/1538-4357/ab12ce}

\bibitem[{{Lada} {et~al.}(2010){Lada}, {Lombardi}, \& {Alves}}]{Lada_2010}
{Lada}, C.~J., {Lombardi}, M., \& {Alves}, J.~F. 2010, \apj, 724, 687,
  \dodoi{10.1088/0004-637X/724/1/687}

\bibitem[{{Madura} {et~al.}(2015){Madura}, {Clementel}, {Gull}, {Kruip}, \&
  {Paardekooper}}]{Madura_2015}
{Madura}, T.~I., {Clementel}, N., {Gull}, T.~R., {Kruip}, C.~J.~H., \&
  {Paardekooper}, J.~P. 2015, \mnras, 449, 3780, \dodoi{10.1093/mnras/stv422}

\bibitem[{{McKee} {et~al.}(2010){McKee}, {Li}, \& {Klein}}]{McKee_2010}
{McKee}, C.~F., {Li}, P.~S., \& {Klein}, R.~I. 2010, \apj, 720, 1612,
  \dodoi{10.1088/0004-637X/720/2/1612}

\bibitem[{{McKee} \& {Ostriker}(2007)}]{mckee_2007}
{McKee}, C.~F., \& {Ostriker}, E.~C. 2007, \araa, 45, 565,
  \dodoi{10.1146/annurev.astro.45.051806.110602}

\bibitem[{{Miville-Desch{\^e}nes} {et~al.}(2017){Miville-Desch{\^e}nes},
  {Murray}, \& {Lee}}]{Miville-Deschenes_2017}
{Miville-Desch{\^e}nes}, M.-A., {Murray}, N., \& {Lee}, E.~J. 2017, \apj, 834,
  57, \dodoi{10.3847/1538-4357/834/1/57}

\bibitem[{{Molinari} {et~al.}(2010){Molinari}, {Swinyard}, {Bally}, {Barlow},
  {Bernard}, {Martin}, {Moore}, {Noriega-Crespo}, {Plume}, {Testi}, {Zavagno},
  {Abergel}, {Ali}, {Anderson}, {Andr{\'e}}, {Baluteau}, {Battersby},
  {Beltr{\'a}n}, {Benedettini}, {Billot}, {Blommaert}, {Bontemps}, {Boulanger},
  {Brand}, {Brunt}, {Burton}, {Calzoletti}, {Carey}, {Caselli}, {Cesaroni},
  {Cernicharo}, {Chakrabarti}, {Chrysostomou}, {Cohen}, {Compiegne}, {de
  Bernardis}, {de Gasperis}, {di Giorgio}, {Elia}, {Faustini}, {Flagey},
  {Fukui}, {Fuller}, {Ganga}, {Garcia-Lario}, {Glenn}, {Goldsmith}, {Griffin},
  {Hoare}, {Huang}, {Ikhenaode}, {Joblin}, {Joncas}, {Juvela}, {Kirk},
  {Lagache}, {Li}, {Lim}, {Lord}, {Marengo}, {Marshall}, {Masi}, {Massi},
  {Matsuura}, {Minier}, {Miville-Desch{\^e}nes}, {Montier}, {Morgan}, {Motte},
  {Mottram}, {M{\"u}ller}, {Natoli}, {Neves}, {Olmi}, {Paladini}, {Paradis},
  {Parsons}, {Peretto}, {Pestalozzi}, {Pezzuto}, {Piacentini}, {Piazzo},
  {Polychroni}, {Pomar{\`e}s}, {Popescu}, {Reach}, {Ristorcelli}, {Robitaille},
  {Robitaille}, {Rod{\'o}n}, {Roy}, {Royer}, {Russeil}, {Saraceno}, {Sauvage},
  {Schilke}, {Schisano}, {Schneider}, {Schuller}, {Schulz}, {Sibthorpe},
  {Smith}, {Smith}, {Spinoglio}, {Stamatellos}, {Strafella}, {Stringfellow},
  {Sturm}, {Taylor}, {Thompson}, {Traficante}, {Tuffs}, {Umana}, {Valenziano},
  {Vavrek}, {Veneziani}, {Viti}, {Waelkens}, {Ward-Thompson}, {White},
  {Wilcock}, {Wyrowski}, {Yorke}, \& {Zhang}}]{Molinari_2010}
{Molinari}, S., {Swinyard}, B., {Bally}, J., {et~al.} 2010, \aap, 518, L100,
  \dodoi{10.1051/0004-6361/201014659}

\bibitem[{{Naiman} {et~al.}(2017){Naiman}, {Borkiewicz}, \&
  {Christensen}}]{Naiman_2017}
{Naiman}, J.~P., {Borkiewicz}, K., \& {Christensen}, A.~J. 2017, \pasp, 129,
  058008, \dodoi{10.1088/1538-3873/aa51b3}

\bibitem[{{Orlando} {et~al.}(2019){Orlando}, {Pillitteri}, {Bocchino},
  {Daricello}, \& {Leonardi}}]{Orlando_2019}
{Orlando}, S., {Pillitteri}, I., {Bocchino}, F., {Daricello}, L., \&
  {Leonardi}, L. 2019, Research Notes of the American Astronomical Society, 11,
  176

\bibitem[{{Padoan} \& {Nordlund}(2002)}]{Padoan_2002}
{Padoan}, P., \& {Nordlund}, {\r{A}}. 2002, \apj, 576, 870,
  \dodoi{10.1086/341790}

\bibitem[{{Padoan} {et~al.}(1997){Padoan}, {Nordlund}, \&
  {Jones}}]{Padoan_1997}
{Padoan}, P., {Nordlund}, A., \& {Jones}, B. J.~T. 1997, \mnras, 288, 145,
  \dodoi{10.1093/mnras/288.1.145}

\bibitem[{{Peng} {et~al.}(2010){Peng}, {Lilly}, {Kova{\v{c}}}, {Bolzonella},
  {Pozzetti}, {Renzini}, {Zamorani}, {Ilbert}, {Knobel}, {Iovino}, {Maier},
  {Cucciati}, {Tasca}, {Carollo}, {Silverman}, {Kampczyk}, {de Ravel},
  {Sanders}, {Scoville}, {Contini}, {Mainieri}, {Scodeggio}, {Kneib}, {Le
  F{\`e}vre}, {Bardelli}, {Bongiorno}, {Caputi}, {Coppa}, {de la Torre},
  {Franzetti}, {Garilli}, {Lamareille}, {Le Borgne}, {Le Brun}, {Mignoli},
  {Perez Montero}, {Pello}, {Ricciardelli}, {Tanaka}, {Tresse}, {Vergani},
  {Welikala}, {Zucca}, {Oesch}, {Abbas}, {Barnes}, {Bordoloi}, {Bottini},
  {Cappi}, {Cassata}, {Cimatti}, {Fumana}, {Hasinger}, {Koekemoer},
  {Leauthaud}, {Maccagni}, {Marinoni}, {McCracken}, {Memeo}, {Meneux}, {Nair},
  {Porciani}, {Presotto}, \& {Scaramella}}]{peng_2010}
{Peng}, Y.-j., {Lilly}, S.~J., {Kova{\v{c}}}, K., {et~al.} 2010, \apj, 721,
  193, \dodoi{10.1088/0004-637X/721/1/193}

\bibitem[{{Press} \& {Schechter}(1974)}]{press_1974}
{Press}, W.~H., \& {Schechter}, P. 1974, \apj, 187, 425, \dodoi{10.1086/152650}

\bibitem[{{Punzo} {et~al.}(2015){Punzo}, {van der Hulst}, {Roerdink},
  {Oosterloo}, {Ramatsoku}, \& {Verheijen}}]{Punzo_2015}
{Punzo}, D., {van der Hulst}, J.~M., {Roerdink}, J.~B.~T.~M., {et~al.} 2015,
  Astronomy and Computing, 12, 86, \dodoi{10.1016/j.ascom.2015.05.004}

\bibitem[{{Rice} {et~al.}(2016){Rice}, {Goodman}, {Bergin}, {Beaumont}, \&
  {Dame}}]{Rice_2016}
{Rice}, T.~S., {Goodman}, A.~A., {Bergin}, E.~A., {Beaumont}, C., \& {Dame},
  T.~M. 2016, \apj, 822, 52, \dodoi{10.3847/0004-637X/822/1/52}

\bibitem[{{Robertson} \& {Goldreich}(2018)}]{Robertson_2018}
{Robertson}, B., \& {Goldreich}, P. 2018, \apj, 854, 88,
  \dodoi{10.3847/1538-4357/aaa89e}

\bibitem[{{Schmidt} {et~al.}(2009){Schmidt}, {Federrath}, {Hupp}, {Kern}, \&
  {Niemeyer}}]{Schmidt_2009}
{Schmidt}, W., {Federrath}, C., {Hupp}, M., {Kern}, S., \& {Niemeyer}, J.~C.
  2009, \aap, 494, 127, \dodoi{10.1051/0004-6361:200809967}

\bibitem[{{Smith} {et~al.}(2014){Smith}, {Glover}, \& {Klessen}}]{Smith_2014}
{Smith}, R.~J., {Glover}, S. C.~O., \& {Klessen}, R.~S. 2014, \mnras, 445,
  2900, \dodoi{10.1093/mnras/stu1915}

\bibitem[{{Torniamenti} {et~al.}(2021){Torniamenti}, {Pasquato}, {Di Cintio},
  {Ballone}, {Iorio}, \& {Mapelli}}]{torniamenti_2021}
{Torniamenti}, S., {Pasquato}, M., {Di Cintio}, P., {et~al.} 2021, arXiv
  e-prints, arXiv:2106.00684.
\newblock \doarXiv{2106.00684}

\bibitem[{Toro(1997)}]{toro_1997}
Toro, E.~F. 1997, Riemann {Solvers} and {Numerical} {Methods} for {Fluid}
  {Dynamics}, 1st edn. (Springer).
\newblock \url{https://link.springer.com/book/10.1007%2F978-3-662-03490-3}

\bibitem[{{Truelove} {et~al.}(1997){Truelove}, {Klein}, {McKee}, {Holliman},
  {Howell}, \& {Greenough}}]{Truelove_1997}
{Truelove}, J.~K., {Klein}, R.~I., {McKee}, C.~F., {et~al.} 1997, \apjl, 489,
  L179, \dodoi{10.1086/310975}

\bibitem[{{Turk} {et~al.}(2011){Turk}, {Smith}, {Oishi}, {Skory}, {Skillman},
  {Abel}, \& {Norman}}]{Turk2011}
{Turk}, M.~J., {Smith}, B.~D., {Oishi}, J.~S., {et~al.} 2011, \apjs, 192, 9,
  \dodoi{10.1088/0067-0049/192/1/9}

\bibitem[{{Vazquez-Semadeni}(1994)}]{Vazquez-Semadeni_1994}
{Vazquez-Semadeni}, E. 1994, \apj, 423, 681, \dodoi{10.1086/173847}

\bibitem[{{Vogt} \& {Shingles}(2013)}]{Vogt_2013}
{Vogt}, F.~P.~A., \& {Shingles}, L.~J. 2013, Astrophysics and Space Science,
  347, 47

\bibitem[{{Wang} \& {Abel}(2009)}]{WangAndAbel_2009}
{Wang}, P., \& {Abel}, T. 2009, \apj, 696, 96,
  \dodoi{10.1088/0004-637X/696/1/96}

\bibitem[{{Williams} {et~al.}(2000){Williams}, {Blitz}, \&
  {McKee}}]{Williams_2000}
{Williams}, J.~P., {Blitz}, L., \& {McKee}, C.~F. 2000, in Protostars and
  Planets IV, ed. V.~{Mannings}, A.~P. {Boss}, \& S.~S. {Russell}, 97.
\newblock \doarXiv{astro-ph/9902246}

\bibitem[{{Winkel} {et~al.}(2021){Winkel}, {Pasquali}, {Kraljic}, {Smith},
  {Gallazzi}, \& {Jackson}}]{winkel_2021}
{Winkel}, N., {Pasquali}, A., {Kraljic}, K., {et~al.} 2021, arXiv e-prints,
  arXiv:2105.13368.
\newblock \doarXiv{2105.13368}

\bibitem[{{Zel'Dovich}(1970)}]{zeldovich_1970}
{Zel'Dovich}, Y.~B. 1970, \aap, 500, 13

\end{thebibliography}

\end{document}